%% file: main.tex
\documentclass[journal]{IEEEtran}

\ifCLASSINFOpdf
\else
\fi

\usepackage{cite}
\usepackage{multirow}
\usepackage{color}
\usepackage{hyperref}
\usepackage{algorithmic}
\usepackage{graphicx}
\usepackage{listings}
\usepackage{xspace}
\usepackage[table]{xcolor}
\usepackage{multirow}
\usepackage{xspace}
\usepackage{subfigure}
\usepackage{listings}
\usepackage{textcomp}
\usepackage{stfloats}
\usepackage[ruled,linesnumbered,noend]{algorithm2e}
\usepackage{booktabs}
\usepackage{url}
\usepackage{hyperref}
\usepackage{enumitem}
\usepackage[cmex10]{amsmath}
\usepackage{amssymb}

\DeclareRobustCommand{\perthousand}{%
  \ifmmode
    \text{\textperthousand}%
  \else
    \textperthousand
  \fi}

\lstdefinestyle{customstyle}{
  breaklines=true,
  frame=single,
  basicstyle=\linespread{1.15}\small,
  numbers=left,
  numberstyle=\small,
  columns=fullflexible,
  showlines=true,
  mathescape=true,
  escapechar={^},
  linewidth=1.04\columnwidth,
  xleftmargin=0.01\columnwidth
}

\begin{document}

\title{MPOCryptoML: Multi-Pattern based Off-Chain Crypto Money Laundering Detection}

\author{Yasaman~Samadi,~Hai~Dong,~\IEEEmembership{Senior Member,~IEEE,}~Xiaoyu~Xia%
\thanks{All authors are with the School of Computing Technologies, RMIT University, Melbourne, VIC, Australia.}
\thanks{Manuscript received July 10, 2025.}
}

\markboth{IEEE Transactions on Information Forensics and Security}{Samadi \MakeLowercase{\textit{et al.}}: MPOCryptoML: Multi-Pattern based Off-Chain Crypto Money Laundering Detection}

\maketitle

\begin{abstract}
Recent advancements in money laundering detection have demonstrated the potential of using graph neural networks to capture laundering patterns accurately. However, existing models are not explicitly designed to detect the diverse patterns of off-chain cryptocurrency money laundering. Neglecting any laundering pattern introduces critical detection gaps, as each pattern reflects unique transactional structures that facilitate the obfuscation of illicit fund origins and movements. Failure to account for these patterns may result in under-detection or omission of specific laundering activities, diminishing model accuracy and allowing schemes to bypass detection. 
To address this gap, we propose the \textit{MPOCryptoML} model to effectively detect multiple laundering patterns in cryptocurrency transactions. MPOCryptoML includes the development of a multi-source Personalized PageRank algorithm to identify random laundering patterns. Additionally, we introduce two novel algorithms by analyzing the timestamp and weight of transactions in high-volume financial networks to detect various money laundering structures, including fan-in, fan-out, bipartite, gather-scatter, and stack patterns. We further examine correlations between these patterns using a logistic regression model. 
An anomaly score function integrates results from each module to rank accounts by anomaly score, systematically identifying high-risk accounts. Extensive experiments on public datasets including Elliptic++, Ethereum fraud detection, and Wormhole transaction datasets validate the efficacy and efficiency of MPOCryptoML. Results show consistent performance gains, with improvements up to 9.13\% in precision, up to 10.16\% in recall, up to 7.63\% in F1-score, and up to 10.19\% in accuracy.
\end{abstract}


\IEEEpeerreviewmaketitle

\input{Introduction}

\input{RelatedWorks}
\input{Priliminaries}

\input{ProblemDefinition}
\input{Model}

\input{ExperimentsandResults}
\input{ApplicationScenarios}
\input{Conclusion}

\section*{Acknowledgment}
This work was fully supported by the CloudTech RMIT Green Bitcoin Joint Research Program.

\bibliographystyle{IEEEtran}
\bibliography{reference_initial}

\ifCLASSOPTIONcaptionsoff
  \newpage
\fi




\vspace{-20mm}
\begin{IEEEbiography}[{\includegraphics[width=1in,height=1.25in,clip,keepaspectratio]{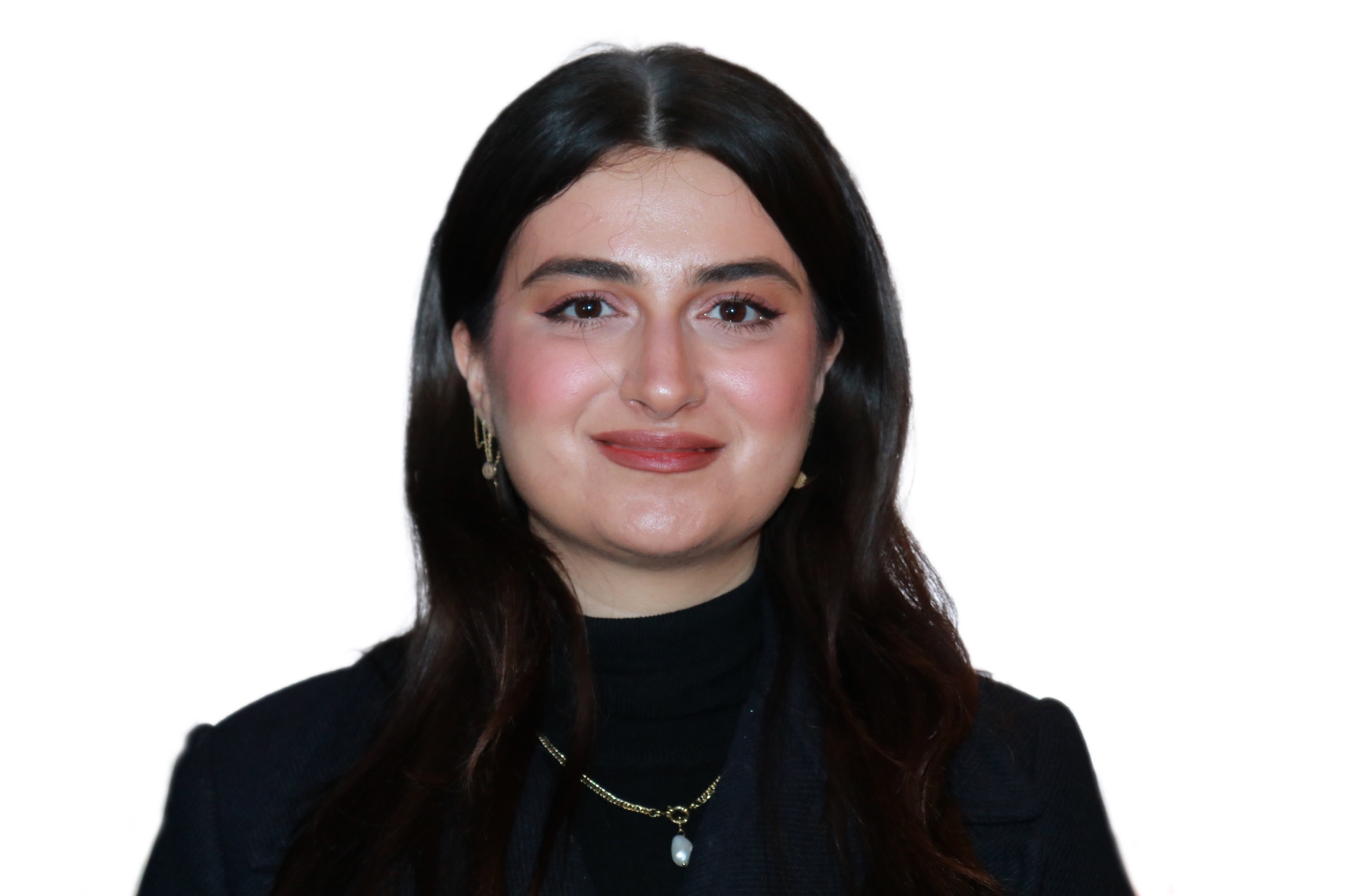}}]{Yasaman Samadi}
Yasaman Samadi is a Ph.D. Candidate at RMIT University, a cybersecurity researcher, and a former quantum computer engineer. Her research focuses on cybersecurity, blockchain fraud detection, and machine learning in financial systems.
\end{IEEEbiography}
\vspace{-20mm}

\begin{IEEEbiography}[{\includegraphics[width=1in,height=1.25in,clip,keepaspectratio]{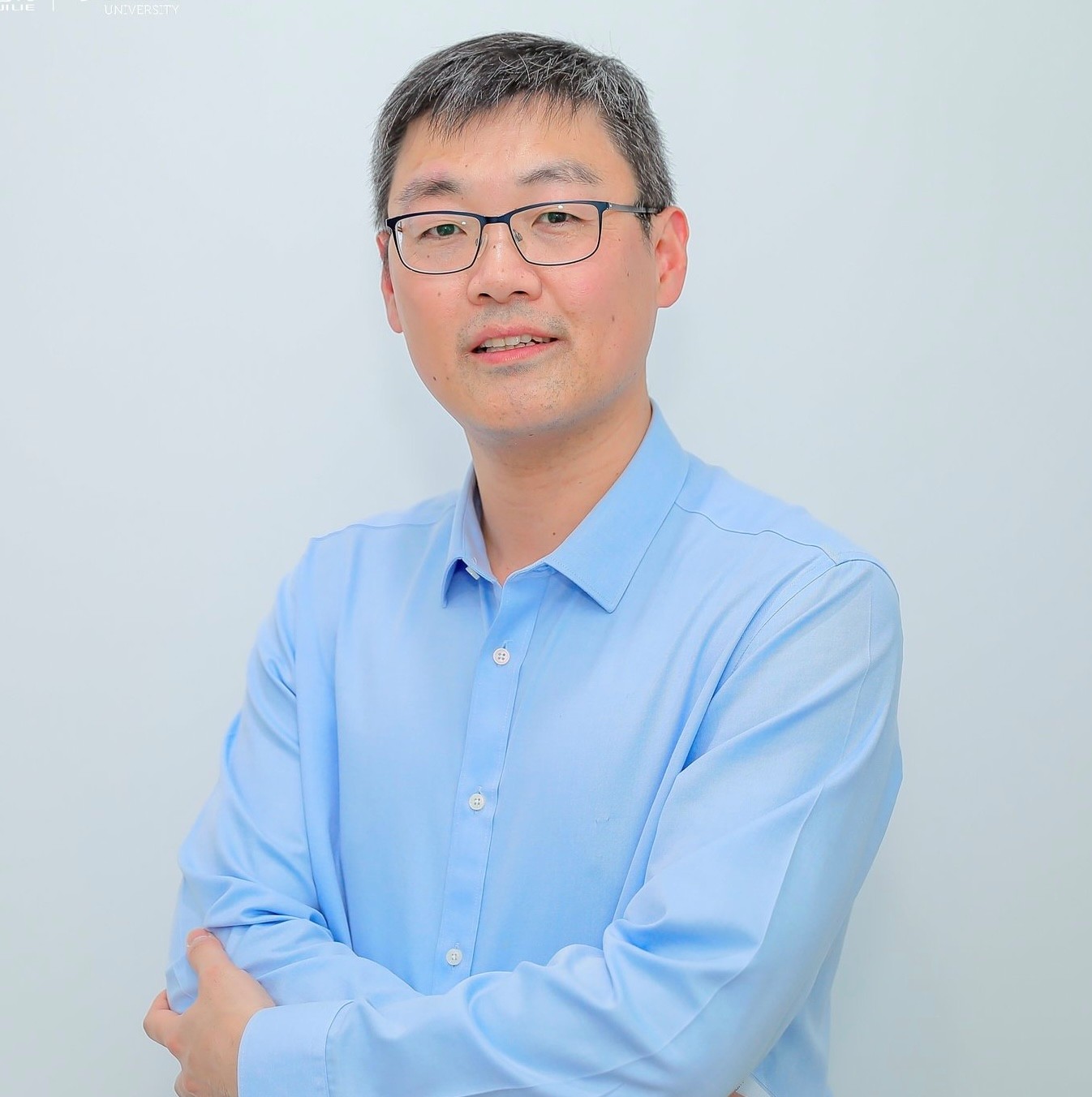}}]{Hai Dong}
(Senior Member, IEEE) 
received a Ph.D. degree from Curtin University, Perth, Australia, and a Bachelor's degree from Northeastern University, Shenyang, China. He is currently a Senior Lecturer at the School of Computing Technologies, RMIT University, Melbourne, Australia. He also serves as the Chair for the IEEE Computational Intelligence Society Task Force on Deep Edge Intelligence. His primary research interests include Cloud/Edge Computing, Edge Intelligence, Blockchain, Cyber Security, and AI Security. His publications appeared in TOSEM, TIFS, TMC, TSC, TSE, AAAI, ACM MM, ASE, ICML, etc. 
\end{IEEEbiography}
\vspace{-20mm}

\begin{IEEEbiography}[{\includegraphics[width=1in,height=1.25in,clip,keepaspectratio]{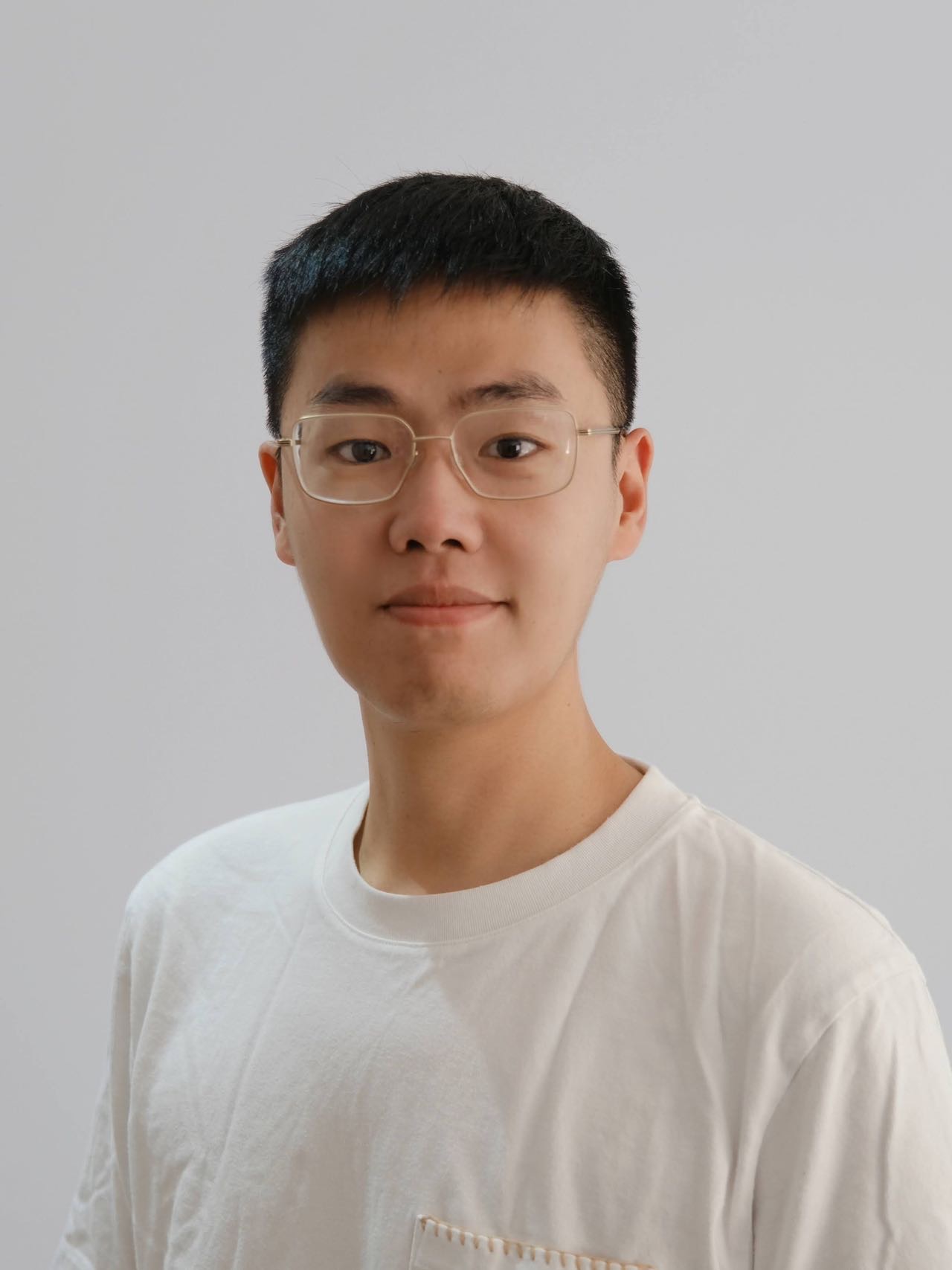}}]{Xiaoyu Xia}
received his PhD degree with the Alfred Deakin Medal from Deakin University, Australia. Currently, he is a lecturer at RMIT University, Australia. He has published over 60 peer-reviewed papers in international journals and conferences, including IEEE S\&P, ACM WWW, ACM SIGIR, IEEE TPDS, IEEE TMC, IEEE JSAC, IEEE TSC, etc. He also serves as an Associate Editor for IEEE TDSC and a Review Board Member of the IEEE TPDS. His research interests include system privacy and security, AI privacy, parallel and distributed computing, and sustainable computing. More details about his research can be found at https://xiaoyushawxia.github.io/homepage/.
\end{IEEEbiography}

\vspace{-20mm}

\end{document}

%% file: Introduction.tex
\section{Introduction}

\IEEEPARstart{C}{rypto} money laundering involves disguising the origins of illegally obtained cryptocurrency to make it appear legitimate ~\cite{yang2023anti}. This process leverages complex transaction chains and obfuscation strategies, often exploiting the decentralized, borderless, and pseudonymous nature of blockchain technologies ~\cite{unitednations}. As cryptocurrency adoption accelerates across financial ecosystems, so does its abuse for illicit purposes, including money laundering, fraud, and sanctions evasion. The global financial and legal implications are profound, necessitating advanced approaches to track and disrupt these activities at scale. A typical example of a crypto money laundering process is illustrated in Figure~\ref{fig:crypto_patterns}, where layered transactions obscure the flow from source to destination addresses.

\begin{figure}[h!]
\vspace{-2mm}
 \begin{center}
        \leavevmode

        \includegraphics[width=0.43\textwidth,keepaspectratio]{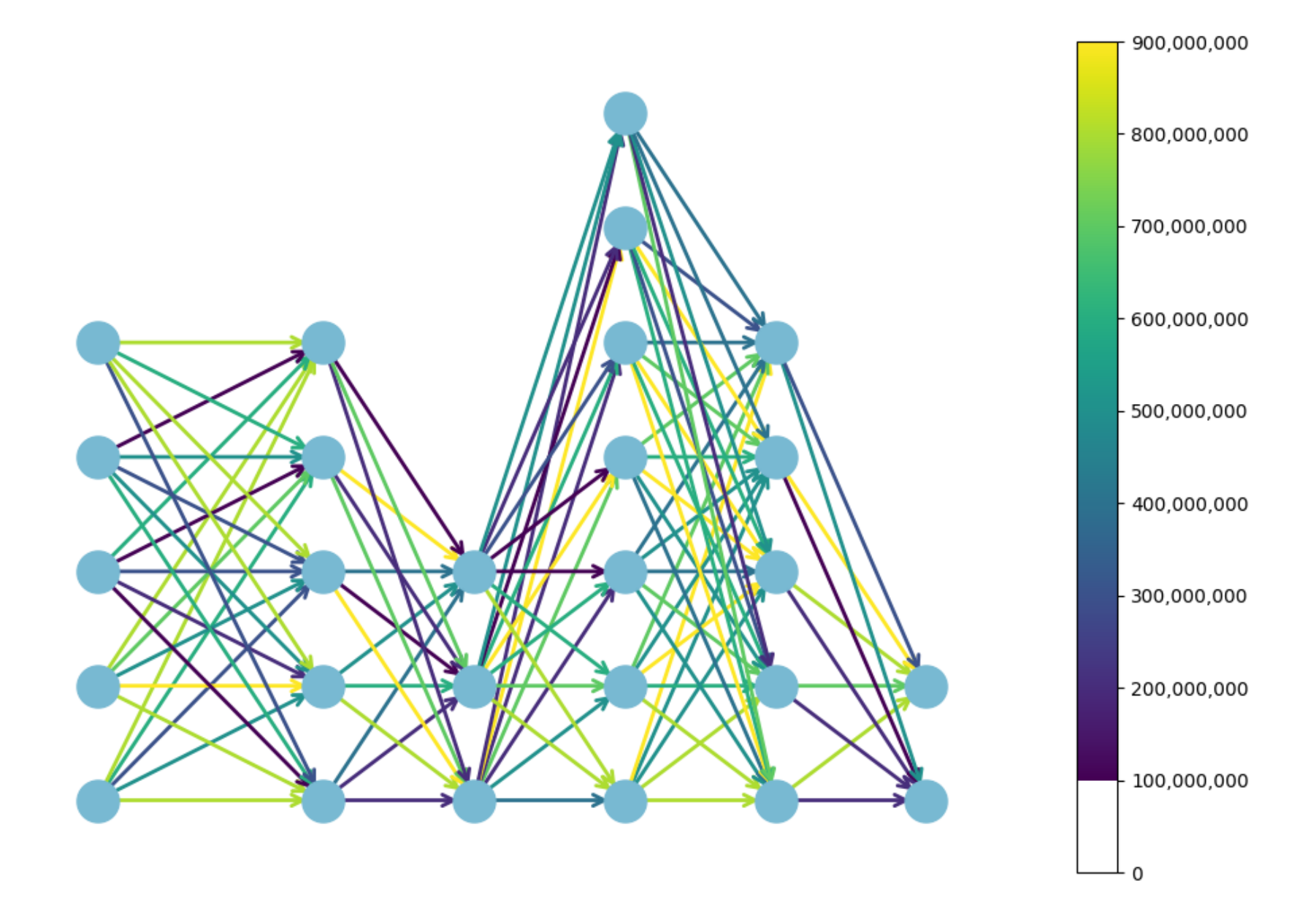}
        \vspace{-5mm}
        \caption{An illustration of cryptocurrency money laundering, which is a series of digital wallet transactions resulting in a complex, multi-pattern subgraph. Sources on the left are used to initiate the money laundering process, which is then carried out through a series of digital wallet stages in the middle to conceal the money's origin and destinations on the right. The amount of money transferred is indicated by the color of the edge.}
        \label{fig:cryptoml_example}
      \end{center}

    \end{figure}

While crypto laundering retains the classical stages of placement, layering, and integration, it departs significantly from traditional bank-based laundering in both execution and traceability. Traditional anti money laundering (AML) frameworks rely heavily on centralized oversight, customer verification, and transaction monitoring, typically confined to one jurisdiction. In contrast, cryptocurrency networks are decentralized, span multiple regulatory zones, and operate pseudonymously \cite{sun2022monlad}. Furthermore, crypto transactions can be automated and executed at scale using scripts and bots, with minimal barriers to entry. These features fundamentally undermine the effectiveness of conventional banking AML mechanisms, rendering many legacy solutions obsolete in crypto contexts.

Crypto transactions can occur both \textit{``on-chain"} (visible and verifiable on the blockchain ledger) and \textit{``off-chain"} (occurring outside the blockchain, often through centralized exchanges, peer-to-peer platforms, or fiat conversion mechanisms). While ``on-chain" activities provide transparency and immutability, ``off-chain" transactions remain opaque, fragmented, and jurisdictionally complex. ``Off-chain" laundering lacks standardized monitoring, making it harder to detect than its ``on-chain" counterpart. Due to these unique characteristics, ``off-chain" laundering often becomes the gateway for illicit crypto assets to re-enter the legitimate financial system.

Existing crypto AML tools primarily target ``on-chain" activities using blockchain analysis, anomaly detection, and graph-based learning models~\cite{AUSTRAC, Chainanalysis_2024}.   
However, these methods struggle with obfuscation techniques, lack visibility into centralized platforms, and are largely ineffective against laundering strategies that span ``off-chain" domains. Conventional models often depend on static laundering patterns or transaction thresholds and fail to adapt to evolving laundering tactics, especially in environments without direct blockchain traceability. Moreover, most prior works focus on wallet security~\cite{kolachala2021sok}, protocol-level vulnerabilities, or generalized fraud detection without addressing the nuanced nature of ``off-chain" laundering.

To address these challenges, we propose \textbf{MPOCryptoML}, a novel ``off-chain" money laundering detection framework. Our approach models laundering behaviors as multi-source heterogeneous graphs and applies a multi-stage detection architecture. It begins by using a multi-source Personalized PageRank (PPR) algorithm to trace random laundering flows. Then, it applies temporal, structural, and transactional features captured via a Timestamp algorithm, a Weight algorithm, and a Logistic Regression model to identify suspicious patterns such as fan-in/fan-out, bipartite, stack, and gather-scatter. Lastly, an anomaly scoring function aggregates and ranks entities by laundering likelihood.

This paper makes the following contributions:
\begin{itemize}
    \item We present the first end-to-end method, MPOCryptoML, tailored for identifying multiple laundering patterns in ``off-chain" cryptocurrency laundering, bridging a critical research gap.
    \item We develop a multi-source Personalized PageRank (PPR) algorithm that captures hidden and random laundering paths in cross-platform transaction graphs.
    \item We introduce a multi-pattern laundering detection algorithm capable of identifying complex transaction patterns such as fan-in/fan-out, bipartite, gather-scatter, and stack formations.
    \item We design a novel anomaly score function for ranking laundering suspects based on multiple behavioral and topological features in graph space.
    \item We demonstrate the scalability and effectiveness of our approach on real-world datasets from Ethereum fraud detection, Wormhole, and Elliptic++, achieving significant improvements of up to 9.13\% in precision, 10.16\% in recall, 7.63\% in F1-score, and 10.19\% in accuracy, compared to seven state-of-the-art baselines. 
\end{itemize}


\noindent The rest of this paper is structured as follows: 
Section \ref{sec:related_work} provides an extensive review of related works, while Section \ref{sec:priliminaries} covers preliminaries on ``off-chain" crypto money laundering workflow and patterns. 
Section \ref{sec:problem_definition} introduces the problem formulation, followed by Section \ref{sec:method}, which details \textit{MPOCryptoML} architecture. 
Section \ref{sec:experiments} presents the experiments, results, and an ablation study to evaluate the model's performance. 
Section \ref{sec:applications} discusses the application scenarios. Finally, Section \ref{sec:conclusion} concludes the paper with key insights and future directions.

%% file: RelatedWorks.tex
\section{Related Work}
\label{sec:related_work}

In this section, we review various methods for detecting anomalies and outliers, as well as identifying patterns of money laundering.

\noindent \textbf{Anomaly Detection Approaches:} Anomaly detection aims to identify data instances that deviate significantly from established normal patterns. Given the scarcity of labeled data in real-world scenarios, unsupervised and semi-supervised methods are widely employed. Unsupervised techniques encompass a range of strategies, each based on different assumptions about the nature of anomalies. Isolation-based methods (e.g., Isolation Forest \cite{liu2012isolation}) detect anomalies by recursively partitioning data and isolating points that require fewer splits, assuming that anomalies are few and different. Density-based methods (e.g., Local Outlier Factor \cite{breunig2000lof}) consider instances in sparse regions as anomalies, relying on the assumption that normal points lie in dense clusters. Distance-based methods (e.g., Deep SVDD \cite{ruff2018deep}) measure how far a point deviates from others or a learned center, identifying distant points as anomalies. Probability-based approaches (e.g., ECOD \cite{li2022ecod}) model the underlying data distribution and flag instances with low probability density as outliers. Reconstruction-based methods (e.g., Deep Autoencoding Gaussian Mixture Model \cite{zong2018deep}) train models to reconstruct normal instances accurately, and instances with high reconstruction error are treated as anomalies. Despite their flexibility, these unsupervised methods often suffer from high false positive rates due to the absence of ground truth and variability in real-world data.

To address these limitations, semi-supervised methods have emerged as a promising alternative by incorporating limited labeled data to guide the anomaly detection process. For example, positive-unlabeled (PU) learning \cite{ju2020pumad, mu2021positive} assumes that labeled anomalies are rare and attempts to identify outliers from a large pool of unlabeled data, though it often relies on assumptions (e.g., uniformity of anomalies) that may not hold in practice. GAN-based techniques \cite{tian2022anomaly, zong2022peripheral} learn to distinguish real data from data generated by a generative adversarial network, with anomalies detected based on the discrepancy. Notably, PIAWAL \cite{zong2022peripheral} enhances this by introducing weighted adversarial learning to amplify anomalous signal strength. Other approaches, such as MACE \cite{chen2024learning}, utilize spectral representations of transaction behavior, enabling efficient and robust anomaly detection through frequency-domain feature extraction.


\noindent \textbf{Anti money Laundering approaches (AML):} Machine learning algorithms are also employed for the detection of money laundering activities. 
In \cite{tang2005developing}, Support Vector Machines (SVM) were utilized to process large datasets, achieving higher accuracy. 
The authors of \cite{michalak2011graph} applied fuzzy matching to identify subgraphs containing suspicious accounts. 
In \cite{lv2008rbf}, the authors determined the involvement of capital flow in ML activities using Radial Basis Function (RBF) neural networks calculated over time. 
Paula et al. \cite{paula2016deep} demonstrated some success in using deep neural networks for this purpose. 
However, these algorithms generally detect ML activities in a supervised manner, which can be problematic due to imbalanced labels and limited adaptability. 
Moreover, ML detection is often an adversarial task, and deep models may lack robustness when facing adversarial attacks.

Graph-based models are particularly effective in detecting complex and non-obvious laundering patterns by capturing structural and temporal relationships between entities. These models typically represent entities (e.g., accounts, addresses) as nodes and interactions (e.g., transactions, fund transfers) as edges, enabling the detection of anomalous substructures within the graph. FlowScope \cite{li2020flowscope} models financial transactions as directed graphs and applies flow tracking to trace the movement of funds across accounts, identifying suspicious routing behaviors indicative of layering stages in money laundering. DiGA \cite{li2023diga} leverages graph mining techniques to extract subgraphs with characteristics typical of anti-money laundering (AML) scenarios, such as fan-in/fan-out or circular patterns, which are often manually curated in traditional systems. AMAP \cite{chai2023towards} utilizes attributed heterogeneous graphs to detect illicit sub-networks by learning node embeddings that reflect both structural roles and attribute similarities, enabling the detection of laundering patterns involving identity obfuscation or account reuse. MoNLAD \cite{sun2022monlad} integrates real-time transaction streams into a dynamic graph model and applies temporal anomaly detection to flag accounts that suddenly exhibit behaviors inconsistent with their historical profile.

While traditional graph-based AML models demonstrate strong performance in capturing relational structures, they are primarily designed for regulated banking systems and face limitations when applied to decentralized, large-scale cryptocurrency environments. To address this gap, several graph-based methods have been proposed specifically for cryptocurrency anti-money laundering (AML). ComGA \cite{luo2022comga} integrates deep multi-layer perceptrons (MLPs) with graph neural networks (GNNs) to learn node representations that capture both transaction features and structural context in crypto networks. AEGIS \cite{wyatt2022anoddpm} employs adversarial learning to detect anomalies by modeling node behaviors under distributional shifts, but its application is hindered by high computational costs in large transaction graphs. Clustering-based approaches, such as DeepFD \cite{wang2018deep} and semi-GCN \cite{kumagai2021semi}, utilize GNN embeddings to group similar nodes and detect outliers. However, these methods struggle with defining clear cluster boundaries in highly dynamic and noisy crypto networks, often resulting in elevated false positive rates. Despite these advancements, no existing method has demonstrated robust and scalable detection of crypto money laundering patterns in real-world, high-volume blockchain data, highlighting the need for more effective graph-based models tailored to the unique characteristics of decentralized financial systems.

Decentralized transactions in blockchain systems present unique challenges for money laundering detection due to their irreversible execution, lack of centralized oversight, and pseudonymous identities. These properties hinder traditional anomaly detection methods, which often rely on centralized monitoring, identifiable entities, and reversible audit trails. Additionally, standard AML frameworks are typically tailored for regulated environments, where transaction semantics, user identities, and activity histories are accessible assumptions that do not hold in permissionless blockchain ecosystems. Tracing illicit fund flows across decentralized networks requires not only anomaly detection but also pattern recognition across multi-hop, temporally dynamic graphs. Existing models in both anomaly detection and AML are often designed to detect only limited laundering patterns (e.g., fan-out, fan-in, stack, gather-scatter, random walks, or bipartite flows). These models typically encode strong prior assumptions, making them ineffective at generalizing to unseen or evolving laundering strategies in the wild. As a result, they exhibit limited accuracy and high false negative rates, leaving significant detection gaps when applied to complex, multi-pattern, and ``off-chain" laundering behaviors common in real-world crypto environments.

%% file: Priliminaries.tex
\section{Priliminaries}
\label{sec:priliminaries}

Crypto money laundering, driven by blockchain technology, exhibits unique features while sharing similarities with traditional laundering techniques \cite{yang2023anti}, \cite{Chainanalysis}, \cite{AUSTRAC}, \cite{aic.gov.au}, \cite{altman2023realistic}, \cite{li2020flowscope}, \cite{li2023diga}, \cite{sun2022monlad}. The decentralized and pseudonymous nature of blockchain introduces new challenges, such as anonymity of addresses, decentralized networks complicating oversight, and the speed and irreversibility of transactions, which reduce intervention opportunities. Criminals often use mixing services or tumblers to obscure transaction trails, while rapid cross-border transactions and the emergence of DeFi platforms and smart contracts create additional laundering opportunities. Centralized and decentralized exchanges are key for converting illicit funds into other assets, necessitating monitoring of trading and withdrawal behaviors. Effective detection requires advanced methods, including graph-based analysis, machine learning, and blockchain forensics, to process large volumes of data in real time and adapt to evolving laundering tactics. Developing robust cryptoAML systems tailored to these challenges is essential for addressing cryptocurrency laundering effectively.

\subsection{Crypto Money Laundering Workflow}

Cryptocurrency money laundering typically follows three core stages that mirror conventional laundering processes: \textbf{placement}, \textbf{layering}, and \textbf{integration}. In the placement stage, illicit funds are introduced into the cryptocurrency ecosystem by converting fiat currency into digital assets or vice versa. The layering stage involves transferring these assets across multiple accounts, wallets, or platforms often across jurisdictions to obscure the origin of the funds. This can include complex patterns such as rapid trades, mixing services, and cross-asset swaps. Finally, the integration stage entails reintroducing the laundered funds into the legitimate economy, either by converting them back to fiat or using them for legal purchases in digital form~\cite{Chainanalysis}.

Exchange platforms play a pivotal role throughout this cycle, facilitating both the conversion of fiat to cryptocurrency and the disbursement of laundered assets. Digital wallet addresses serve as intermediaries, allowing illicit actors to move tainted funds through multiple hops before arriving at a final deposit address associated with an exchange. Notably, such laundering activity can be confined entirely within a single blockchain ecosystem, underscoring how decentralised platforms can be exploited for illicit financial flows~\cite{Chainanalysis}.

To mitigate these risks, cryptocurrency exchanges and forensic analytics firms actively monitor trading activity and trace the flow of suspicious funds. For instance, Chainalysis provides investigative support by labeling wallet addresses involved in reported suspicious activity, allowing law enforcement and financial institutions to follow the trail of illicit assets~\cite{dupuis2020money, Chainanalysis}. Regulatory frameworks increasingly require platforms to implement address monitoring and reporting mechanisms, though limitations persist depending on the type of cryptocurrency, the exchange infrastructure, and jurisdictional compliance requirements~\cite{aic.gov.au, ministers.treasury.gov.au}. These complexities make detection and prevention of cryptocurrency money laundering a critical challenge for financial crime investigators and compliance professionals alike.

\subsection{Difference Between On-chain and Off-chain Transactions}

Understanding the fundamental distinctions between ``on-chain" and ``off-chain" transactions is essential for analysing the behavior of illicit financial activities in blockchain ecosystems. These two transaction types differ significantly in terms of transparency, scalability, execution mechanisms, and susceptibility to abuse. On-chain transactions are inherently traceable and governed by the blockchain’s consensus protocol, whereas ``off-chain" transactions bypass direct ledger recording, enabling faster and more private exchanges. This difference is particularly consequential in the context of crypto money laundering, where criminals may exploit ``off-chain" mechanisms to evade detection. Table~\ref{tab:onchain_offchain} summarises the key characteristics that differentiate ``on-chain" and ``off-chain" transactions, highlighting their respective implications for forensic analysis and anti-money laundering efforts.

\begin{table*}[h!]
\centering
\renewcommand{\arraystretch}{1.15}
\caption{Major Differences Between On-Chain and Off-Chain Transactions}
\label{tab:onchain_offchain}
\vspace{-2mm}
\begin{tabular}{|c|l|l|}
\hline
\textbf{Aspect} & \textbf{On-Chain Transactions} & \textbf{Off-Chain Transactions} \\
\hline
\textbf{Execution Location} & Performed directly on the blockchain ledger & Occurs outside the blockchain ledger \\
\textbf{Consensus Involvement} & Requires consensus (e.g., PoW, PoS) for validation & Does not require blockchain consensus \\
\textbf{Transparency} & Publicly visible and auditable & Hidden from public view; often private \\
\textbf{Immutability} & Immutable once confirmed on the blockchain & Can be modified or reversed before final settlement \\
\textbf{Scalability} & Limited by block size and network throughput & Highly scalable and faster \\
\textbf{Fees} & Subject to transaction fees (e.g., gas fees) & Typically low or no blockchain fees \\
\textbf{Latency} & Can be delayed due to network congestion & Enables near-instant settlement \\
\textbf{Security Guarantees} & Secured by cryptographic consensus protocols & Relies on trust, escrow, or external enforcement \\
\textbf{Use in Illicit Activities} & Easier to trace using blockchain analytics & Commonly exploited for laundering due to reduced traceability \\
\hline
\end{tabular}
\end{table*}

\subsection{Off-chain Crypto Money Laundering Patterns}

\begin{figure}[h]
 \vspace{-2mm}
      \begin{center}
        \leavevmode
       \includegraphics[width=0.9\linewidth,keepaspectratio]{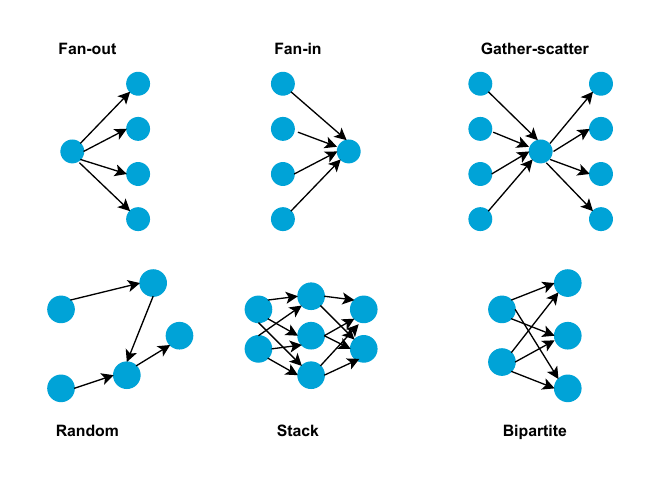}
       \vspace{-7mm}
        \caption{Crypto money laundering patterns \cite{altman2023realistic}}
       \label{fig:crypto_patterns}
      \end{center}
 \vspace{-2mm}
    \end{figure}

Malicious node behaviors in cryptocurrency networks can be modeled as distinct graph problems, which are crucial for addressing money laundering in real-world scenarios and on exchange platforms. These behaviors often manifest as anomalies within groups of nodes in local subgraphs, deviating from typical single-node or individual transaction patterns.
User groups typically form discrete but internally dense subgraphs, which helps address the issue of intrinsic sparsity. Transactions can be represented as weighted directed graphs, where nodes represent accounts or parties, and edges represent asset transfers. The direction of an edge indicates the source and destination of the transfer, while the weight quantifies the value of the transfer.
Analyzing transactions as weighted directed graphs provides a robust framework for understanding asset movement, detecting anomalies, and identifying money laundering patterns in cryptocurrency networks. This approach reveals various patterns of cryptocurrency money laundering, as illustrated in Figure~\ref{fig:crypto_patterns}.

When exploring deeper into the intricacies of crypto money laundering, it is imperative to comprehend the laundering patterns,  that criminals employ to hide their operations. 
\textbf{Fan-in} is a prevalent pattern in which money is sent from several sources to one location, combining money from different sources into one account. 
On the other hand, the \textbf{fan-out} pattern entails a single source sending money to several locations, distributing money among different accounts to make tracing more difficult. 
These strategies are combined in the \textbf{gather-scatter} pattern, which distributes the aggregated amount to several accounts after sending several tiny amounts to one account in the gathering phase (scattering phase). 
The \textbf{bipartite} pattern is another advanced technique that isolates direct contacts and hides account links. 
It consists of two sets of nodes with transactions happening only between nodes of distinct sets. 
Furthermore, the \textbf{random} pattern involves transactions that appear to happen at random, which makes it difficult to detect using conventional analytical techniques. 
Last, the \textbf{stack} pattern is continuously moving money through a sequence of intermediate accounts in a stacked, linear manner, producing a trail of transactions that essentially hides the original source of the money.
The mentioned laundering patterns -fan-in, fan-out, bipartite, gather-scatter, and stack— are suggested by IBM \cite{altman2023realistic}. 
Comprehending these trends is essential to creating effective techniques to identify and stop cryptocurrency money laundering.

%% file: ProblemDefinition.tex
\section{Problem Definition}
\label{sec:problem_definition}

For the crypto money transfers, let $G = (V, E, W, T)$ be a directed graph, with $V$ representing the vertex set. 
The digital wallets' accounts are represented by $V$, the set of edges by $E$, the total amount of money in each account by $W$, and the timestamp at which the edges are added to each node by $T$.
In our case, we refer to the vertex and account interchangeably throughout the study. 

\textbf{\emph {Definition 1}} (\emph{Off-chain crypto money laundering Detection}).
Given a graph of transactions $G = (V, E, W, T)$, the primary objective is to detect and identify multiple ``off-chain" crypto money laundering patterns. 
These patterns are specific subgraphs $G\;' = (V\;', E\;') $, where $V\;' \subseteq V$  and $E\;' \subseteq E$, that exhibit particular structural characteristics indicative of ``off-chain" crypto money laundering activities. 
The detection of these patterns involves analysing the topological features such as fan-in where a vertex $v \in V$ is the number of incoming edges to $v$. 
It is defined as fan-in$(v) = |\{u \in V | (u, v) \in E\}|$.
\begin{equation}   
   \text{fan-in}(v) = {d_i}^-(S) = \Sigma_{{v_k} \in {M_l}_{-1}\wedge(k, v) \in E} e_{kv}
        \label{eqn:fan-in}
\end{equation}

The fan-out of a vertex $v \in V$ is the number of outgoing edges from $v$.
It is defined as fan-out$(v) = |\{w \in V | (v, w) \in E\}|$.
\begin{equation}   
   \text{fan-out}(v) = {d_i}^+(S) = \Sigma_{{v_j} \in {M_l}_{+1}\wedge(v, j) \in E} e_{vj}
        \label{eqn:fan-out}
\end{equation}
The gather-scatter metric for a vertex $v \in V$ combines both the fan-in and fan-out characteristics to evaluate the overall connectivity of $v$.
It is defined as gather-scatter$(v)$ = fan-in$(v)$ + fan-out$(v)$.
\begin{align}
   & \text{gather-scatter}(v) = d_i^-(S) + d_i^+(S) = \notag \\
   & \Biggl( \sum_{v_k \in M_{l-1} \wedge (k, v) \in E} e_{kv} \Biggr) + \notag 
   \Biggl( \sum_{v_j \in M_{l+1} \wedge (v, j) \in E} e_{vj} \Biggr)
   \label{eqn:gather-scatter}
\end{align}

The random refers to the randomness in the occurrence of transactions.
For a vertex $v \in V$, the randomness of edges can be represented by the probability of an edge occurrence as random$(v)$ = Probability $((v, w) \in E)$.
\begin{equation}   
   \text{random}(v) = \frac{\sum_{w \in M_{l+1} \wedge (v, w) \in E} e_{vw}}{|M_{l+1}|}
   \label{eqn:random}
\end{equation}

The stack feature is represented by a directed path of nodes connected in a specific order.
For a path \( P = (v_1, v_2, \dots, v_k) \) of length \( k \), where each \( (v_i, v_{i+1}) \in E \), \( v_i \in M_l \), \( v_{i+1} \in M_{l+1} \), and \( P \) is defined as:
\[
P = (v_1, v_2, \dots, v_k)
\]

A graph is bipartite if its vertex set $V$ can be partitioned into two disjoint sets $V_1$ and $V_2$ such that every edge connects a vertex in $V_1$ to a vertex in $V_2$ and is defined as 
\[
\forall (u, v) \in E, u \in M_l \Rightarrow v \in M_{l+1} \text{ and vice versa}
\]

Table~\ref{tab:notations} presents the notations and their meanings used in algorithms.

\begin{table}[h!]
\renewcommand{\arraystretch}{1.15}
\centering
\caption{Table of notations and meanings}\label{tab:notations}
\resizebox{0.49\textwidth}{!}{
\begin{tabular}{|c|l|}
\hline
\textbf{Notation} & \textbf{Description}  \\
\hline
$u_{i}$  & A node of user account   \\
$M_l$ & Set of nodes at layer $l$ \\
$\alpha$ & Decay factor (the probability that a random walk terminates at a step) \\
$K(s)$      & Total number of random walks for source node  \\
$d(s)$ & Number of out-degrees of the node $s$ \\
$d(u_i)$ & Number of out-degrees of the node $u_i$ \\
${\delta, \epsilon, P_{f}}$ & Parameters of an approximate PPR where, $\delta$ is threshold, $\epsilon$ is error \\ & bound, $p_{f}$ is failure probability \\
$N^m_{s}$ & Number of m-hop out-neighbors of the source node $s$ \\
$r(s, u_i)$ & Residue value of node $u_i$ w.r.t source node $s$ \\
$\sigma(v_{i})$ & Anomaly score\\
$N\theta (v_i)$  & Normalized $\theta$ score for node $v_i$ \\
$N\omega (v_i)$  & Normalized $\omega$ score for node $v_i$ \\
$F_{(\theta, \omega)}(v_i)$ & Set of new features \\
\hline
\end{tabular}}
\end{table}

%% file: Model.tex
\section{MPOCryptoML}
\label{sec:method}

\begin{figure*}[h!]\vspace{-2mm}
      \begin{center}
        \leavevmode
        
        \includegraphics[width=\textwidth, keepaspectratio]{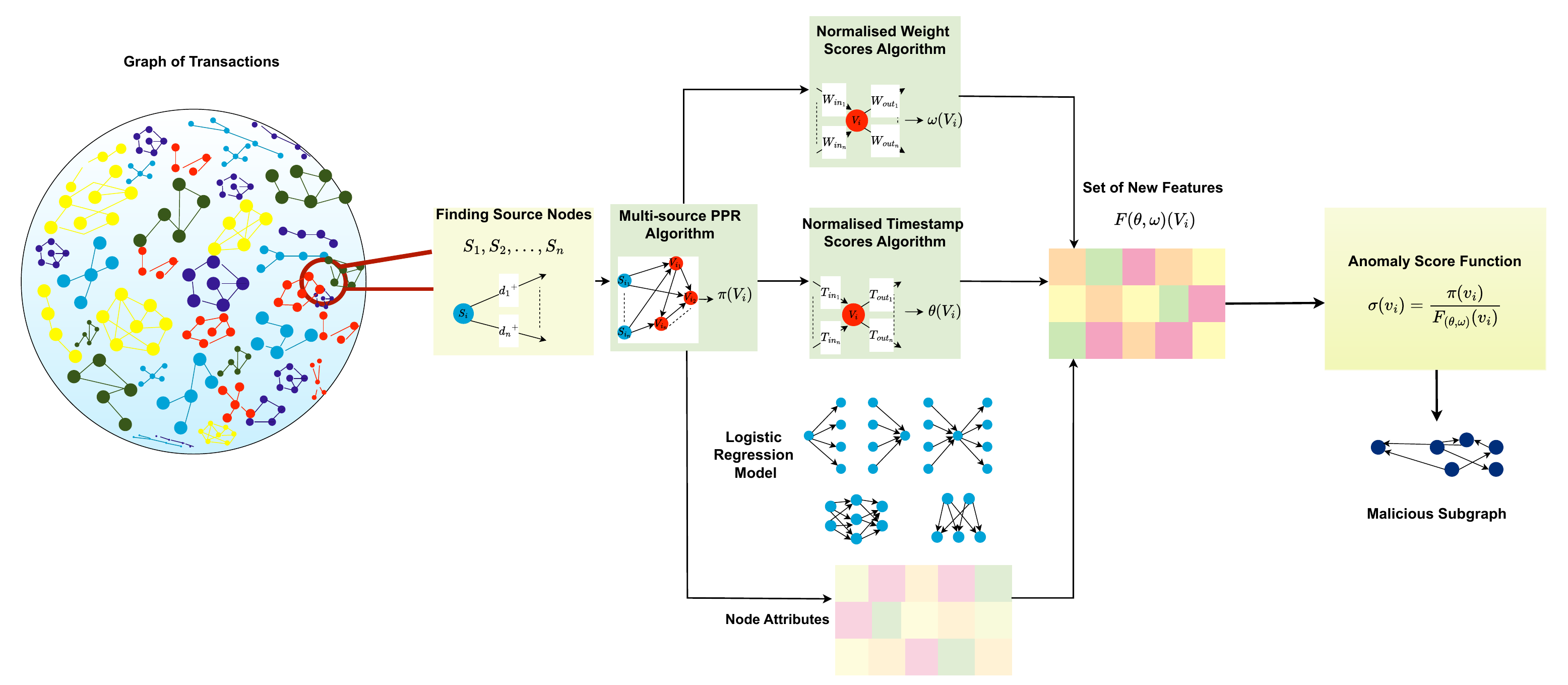}
        \vspace{-7mm}
        \caption{Crypto-AML inference with MPOCryptoML involves several steps: The multi-source Personalised PageRank algorithm first identifies random laundering patterns. Nodes identified in this stage are then evaluated using a normalized timestamp score algorithm and a normalized weight score algorithm. The scores from these components, along with laundering patterns, are input into a logistic regression model to generate new features for each node to detect additional laundering patterns. These PPR scores and new features are then used by an anomaly score function to compute the anomaly score for each node, ultimately identifying those involved in off-chain cryptocurrency money laundering.}
        \label{fig:arch}
      \end{center}
\vspace{-2mm}
    \end{figure*}

This section presents the general framework of the proposed model MPOCryptoML, as shown in Figure~\ref{fig:arch}. 
The primary goal of MPOCryptoML is to identify multiple patterns of ``off-chain" crypto money laundering. To track potential laundering routes in a large transaction graph, the Multi-source PPR algorithm is applied to source nodes \cite{li2023diga}, \cite{wang2017fora}, \cite{luo2019baton}, uncovering random connections and paths between nodes and sources. The PPR scores of each visited node are stored in a tuple called the set of PPR scores (SPS), which is later used in the anomaly score function.
Next, the model evaluates the set of normalized timestamp scores (NTS) and normalized weight scores (NWS) to detect additional laundering patterns. Timestamp scores capture anomalies in transaction timing, while weight scores highlight imbalances in transaction volumes. Normalizing these scores ensures comparability and accurate anomaly detection.
The NTS and NWS values are analyzed using the multi-pattern crypto money laundering detection algorithm, which trains a logistic regression (LR) model to identify laundering patterns. The resulting pattern scores, denoted as $F_{(\theta, \omega)}$, are then input into the Anomaly Score Function.
This comprehensive feature set, including both SPS and $F_{(\theta, \omega)}$, enables the detection of suspicious accounts by identifying various laundering patterns such as fan-in, fan-out, gather-scatter, bipartite, and stack structures, providing a robust method for spotting potential money laundering activities within cryptocurrency transactions.

\subsection{Multi-Source Personalized PageRank Algorithm}

Detecting ``off-chain" cryptocurrency money laundering is critical due to its significant economic and legal implications. To uncover these hidden laundering techniques, identifying random patterns in transaction graphs is essential, as launderers often use complex strategies to obscure illicit payments. The PPR algorithm excels at detecting such random patterns by highlighting potential laundering paths that might otherwise go unnoticed.
PPR ranks nodes based on their relevance to source nodes, focusing on the likelihood of visiting nodes from multiple starting points specifically, the nodes initiating transactions. This probabilistic approach identifies nodes with strong transactional links to the source nodes, assigning them higher relevance scores within dense transactional clusters. By iteratively updating residue and reserve values through random walks, PPR effectively identifies nodes frequently visited, increasing the likelihood of detecting money laundering activities \cite{li2023diga} \cite{luo2019baton}.

The primary objective of using the PPR algorithm is to detect malicious nodes that randomly connect to user accounts (source nodes $s$).
Given the computational expense of performing random walks on large transaction networks, we optimize the process by reducing the number of random walks and employing forward pushes to compute PPR scores more efficiently through a k-hop PPR approach.
To identify \textbf{random} patterns, Algorithm~\ref{alg:ppr} assigns personalised probabilities to multiple sources, representing suspicious nodes within the transaction network. 
Nodes with high PageRank scores are likely involved in laundering activities, as their close connections to the initial suspicious nodes enable the identification of hidden and randomly dispersed laundering patterns across the network.

To address the CryptoAML detection problem, Algorithm~\ref{alg:ppr} is applied to estimate the multi-source Personalized PageRank (PPR) scores across the transaction network. The algorithm is initiated by identifying all source nodes—i.e., addresses that initiate transactions but do not receive any—by locating nodes with outgoing edges but zero in-degree (lines 3–4). These nodes are central to laundering behaviors and are used as seed points for propagation.

Following source identification, Algorithm~\ref{alg:ppr} initializes two vectors: a residual vector $r(s, \cdot)$ and an intermediate score vector $\pi^\circ(s, \cdot)$ for each source node $s$ and its neighbors (lines 5–7). This initialization is necessary to prepare for propagation, where the residual vector reflects the remaining probability mass to be distributed and the score vector accumulates PPR scores.

For each source node $s$, the walk budget $K(s)$ is computed using the Monte Carlo approximation method from~\cite{luo2019baton} (line 9):

\begin{equation}
K(s) = \frac{(2 / 3\epsilon + 2) \cdot d(s) \cdot \log\left(\frac{2}{p_f}\right)}{\epsilon^2 \cdot \alpha(1 - \alpha)}
\end{equation}

where $d(s)$ denotes the out-degree of node $s$, $\epsilon$ is the relative accuracy guarantee, $p_f$ is the failure probability, and $\alpha \in (0,1)$ is the teleport probability. This formulation ensures that highly connected nodes receive a proportionally higher number of walks, increasing statistical robustness in score estimation.

The algorithm’s core propagation stage (lines 10–14) executes a forward push from each node $u_i$ whose residual exceeds a dynamic threshold $\frac{d(u_i)}{\alpha \cdot K(s)}$. When this condition is met, the residual at $u_i$ is distributed to its out-neighbors (line 12), and a portion (proportional to $\alpha$) is added to its own score vector (line 13), followed by resetting the residual at $u_i$ to zero (line 14). This stage reflects a localized score propagation mechanism, ensuring nodes with strong transactional ties accumulate higher scores.

Once no more nodes exceed the residual threshold, the temporary scores are copied to the final score vector $\hat{\pi}(s, \cdot)$ and stored in the result set \texttt{SPS} (lines 15–18). All nodes with nonzero scores are added to the visited node set \texttt{SVN}, which tracks coverage of the transaction subgraph.

To account for remaining residuals, the algorithm proceeds with a Monte Carlo refinement phase (lines 19–24). For each node with residual mass, $r(s, v_i) \cdot K(s)$ random walks are simulated. If a random walk terminates at node $v_i \in N^m_s$, a score increment of $\frac{1}{K(s)}$ is added to $\hat{\pi}(s, v_i)$ (line 24). This reinforces the ranking of nodes frequently reached by random walks, capturing diffusion-based anomalies not propagated in earlier steps.

The final PPR scores (\texttt{SPS}) and visited nodes (\texttt{SVN}) are returned (line 28), completing the estimation procedure. The pseudocode is shown in Algorithm~\ref{alg:ppr} and follows the original PPR formulation with adaptations for multi-source laundering detection~\cite{li2023diga, luo2019baton}.

\RestyleAlgo{ruled} 
\begin{algorithm} [h!]
  \scriptsize
    \caption{Multi-Source PPR Algorithm}\label{alg:ppr}
    \KwIn{Graph $G = (V, E, W, T)$, source node $s$, PPR relative accuracy guarantee $\epsilon$, failure probability $p_f$, proportion allocation $\alpha$, Hop $K$}
    \KwOut{A set of PPR scores (SPS), a set of visited nodes (SVN)}

    $s_{1,2,..,n} \gets 0$\;
    
    \ForEach{$v_i \in V$}{
        \If{in-degree $(v_i) = 0$}{
            Add $v_i$ to the list of source nodes\ \tcp*[r]{Identify source nodes with no incoming edges}
        }
    }

    $r(s, s) \gets 1$; ${\pi}^\circ(s, s) \gets 0$ \tcp*[r]{Initialize residual and score for source node $s$}
    
    \ForEach{$u_i \in N^m_{s}$}{
        $r(s, u_i) \gets 0$, ${\pi}^\circ(s, u_i) \gets 0$\ \tcp*[r]{Initialize neighbors of $s$ with 0 residual and score}
    }

    \ForEach{$s$}{
        $K(s) \gets \frac{(\frac{2}{3}\epsilon + 2) d(s) \log \left(\frac{2}{p_{f}}\right)}{\epsilon^2 \alpha (1 - \alpha)}$ \tcp*[r]{Compute walk count based on accuracy, $\alpha$, and degree}
        
        \While{exists $u_i$ such that $r(s, u_i) > \frac{d(u_i)}{\alpha \cdot K(s)}$ and $d(u_i) > 0$}{
            \ForEach{$v_j$ that is an out-neighbor of $u_i$}{
                $r(s, v_j) \gets r(s, v_j) + (1 - \alpha) \cdot \frac{r(s, u_i)}{d(u_i)}$\ \tcp*[r]{Push residual to out-neighbors}
                ${\pi}^\circ (s, u_i) \gets {\pi}^\circ (s, u_i) + \alpha \cdot r(s, u_i)$\ \tcp*[r]{Update score with $\alpha$ portion}
                $r(s, u_i) \gets 0$\ \tcp*[r]{Reset residual after push}
            }
        }

        \ForEach{$v_i \in N^m_{s}$}{
            $\hat{\pi} (s, v_i) \gets {\pi}^\circ (s, v_i)$\ \tcp*[r]{Copy temporary score to final score}
            Add $\hat{\pi} (s, v_i)$ to $SPS$\ \tcp*[r]{Store score in the result set}
            Add $v_i$ to $SVN$\ \tcp*[r]{Mark node as visited}
        }

        \ForEach{$v_i \in N^m_{s}$ and $r(s, v_i) > 0$}{
            \For{$i = 1$ to $(r(s, v_i) \cdot K(s))$}{
                Conduct a random walk from $v_i$: $(1 - \alpha) \cdot r(s, u_i) \cdot K(s)$\ \tcp*[r]{Simulate diffusion via random walk}
                \If{the random walk terminates at $v_i$ and $v_i \in N^m_{s}$}{
                    $K(u_i) \gets r(s, u_i) \cdot K(s)$\ \tcp*[r]{Update walk budget}
                    $\hat{\pi} (s, v_i) \gets \hat{\pi} (s, v_i) + \frac{1}{K(s)}$\ \tcp*[r]{Increment score based on walk result}
                    Add $\hat{\pi} (s, v_i)$ to $SPS$\ \tcp*[r]{Store updated score}
                    Add $v_i$ to $SVN$\ \tcp*[r]{Add node to visited list}
                }
            }
        }
    }

    \Return $SPS$ and $SVN$\ \tcp*[r]{Return final scores and visited nodes}
\end{algorithm}

\subsection{Multi-Pattern Crypto Money Laundering Algorithm}

Money laundering accounts often exhibit distinctive behavior, such as receiving a large number of transactions within a short period and quickly transferring the funds to other accounts. 
This pattern is indicative of attempts to obscure the origins and destinations of illicit funds. 
Unlike conventional accounts, money laundering accounts tend to show high transactional activity, both in volume and frequency. 
Additionally, these accounts typically transfer nearly the same amount of money in and out, creating a symmetry in transaction amounts. 
This symmetry is a deliberate strategy to maintain the appearance of regular, balanced activity while avoiding detection. 
By closely matching the amounts of incoming and outgoing transactions, these accounts make it difficult to identify suspicious activity based solely on volume discrepancies. Moreover, the high transaction turnover and volume contribute to the high density of these nodes within the transaction graph, complicating their detection. 

Algorithm~\ref{alg:timestampscore} computes the \textit{Normalized Timestamp Score (NTS)} for each node by capturing the temporal asymmetry between its incoming and outgoing transactions. This metric is particularly effective for identifying transient intermediary accounts—commonly used in money laundering—where funds are rapidly cycled through addresses to conceal origin and destination. These accounts typically exhibit minimal temporal separation between incoming and outgoing activity.

The algorithm operates on the set of visited nodes $\text{SVN}$, identified by the Personalized PageRank (PPR) algorithm, along with two auxiliary mappings: the in-degree timestamp set $\text{SITS}$ and the out-degree timestamp set $\text{SOTS}$. For each node $v_i \in \text{SVN}$, the goal is to derive a normalized timestamp deviation score $N_\theta(v_i) \in [0,1]$ representing how temporally unbalanced the node’s transaction behavior is.

The procedure begins by computing the temporal spread of the in-degree timestamps $T_{s_{\text{in}}}(v_i) \in \text{SITS}(v_i)$ using the difference between maximum and minimum timestamps (lines 3–8):
\begin{equation}
    \theta_{\text{in}}(v_i) = \max T_{s_{\text{in}}}(v_i) - \min T_{s_{\text{in}}}(v_i)
\end{equation}

Similarly, it computes the temporal spread of the out-degree timestamps $T_{s_{\text{out}}}(v_i) \in \text{SOTS}(v_i)$ (lines 9–14):
\begin{equation}
    \theta_{\text{out}}(v_i) = \max T_{s_{\text{out}}}(v_i) - \min T_{s_{\text{out}}}(v_i)
\end{equation}

The core measure of temporal irregularity is then calculated as the absolute difference between the two spreads (line 15):
\begin{equation}
    \theta(v_i) = \left| \theta_{\text{out}}(v_i) - \theta_{\text{in}}(v_i) \right|
\end{equation}

This raw difference score $\theta(v_i)$ is stored in a temporary set $\text{STS}$ (line 16) for later normalization. To facilitate comparison across nodes with diverse activity patterns, the algorithm applies min-max normalization over all nodes (lines 18–21):
\begin{equation}
    N_\theta(v_i) = \frac{\theta(v_i) - \min_v \theta(v)}{\max_v \theta(v) - \min_v \theta(v)}
\end{equation}

The resulting score $N_\theta(v_i)$ indicates how temporally asymmetric a node's behavior is. Nodes with low scores are indicative of quick-turnaround behavior—characteristic of laundering intermediaries—while high scores suggest delayed redistribution or asset holding, which are less indicative of illicit flow.

The full pseudocode for this timestamp scoring mechanism is presented in Algorithm~\ref{alg:timestampscore}.

\RestyleAlgo{ruled} 
\begin{algorithm} [h!]
  \scriptsize
    \caption{Normalised Timestamp Score Algorithm} \label{alg:timestampscore}
    \KwIn{A set of visited nodes $SVN$, a set of in-degree timestamps for all nodes $SITS$, a set of out-degree timestamps for all nodes $SOTS$, where $SITS(v_i) \in SITS$ represents the SITS for node $v_i$, and $SOTS(v_i) \in SOTS$ represents the SOTS for node $v_i$}
    \KwOut{A set of normalized timestamp scores $NTS = \{N\theta(v_1), N\theta(v_2), \ldots, N\theta(v_i)\}$}

    \ForEach{$v_i \in SVN$}{ \tcp*[f]{Iterate over each visited node}
    
        \tcp{--- Compute in-degree timestamp range ---}
        $\forall TS_{in} (v_i) \in SITS(v_i)$\ 
        Max $TS_{in} (v_i) \gets -\infty$\ 
        Min $TS_{in} (v_i) \gets +\infty$\; 
        
        \ForEach{$TS_{in}(v_i) \in SITS(v_i)$}{ \tcp*[f]{Iterate over in-degree timestamps}
            \If{$TS_{in}(v_i) >$ Max $TS_{in}(v_i)$}{
                Max $TS_{in}(v_i) \gets TS_{in}(v_i)$\ 
            }
            \If{$TS_{in}(v_i) <$ Min $TS_{in}(v_i)$}{
                Min $TS_{in}(v_i) \gets TS_{in}(v_i)$\
            }
        }

        \tcp{--- Compute out-degree timestamp range ---}
        $\forall TS_{out} (v_i) \in SOTS(v_i)$\
        Max $TS_{out} (v_i) \gets -\infty$\
        Min $TS_{out} (v_i) \gets +\infty$\
        
        \ForEach{$TS_{out}(v_i) \in SOTS(v_i)$}{ \tcp*[f]{Iterate over out-degree timestamps}
            \If{$TS_{out}(v_i) >$ Max $TS_{out}(v_i)$}{
                Max $TS_{out}(v_i) \gets TS_{out}(v_i)$\
            }
            \If{$TS_{out}(v_i) <$ Min $TS_{out}(v_i)$}{
                Min $TS_{out}(v_i) \gets TS_{out}(v_i)$\
            }
        }

        $\theta_{in}(v_i) \gets$ Max $TS_{in}(v_i) -$ Min $TS_{in}(v_i)$\ \tcp*[r]{Compute in-degree timestamp spread}
        $\theta_{out}(v_i) \gets$ Max $TS_{out}(v_i) -$ Min $TS_{out}(v_i)$\ \tcp*[r]{Compute out-degree timestamp spread}
        $\theta(v_i) \gets |\theta_{out}(v_i) - \theta_{in}(v_i)|$\ \tcp*[r]{Absolute difference between out- and in-degree spreads}
        Add $\theta(v_i)$ to $STS$\ \tcp*[r]{Store the unnormalized score}
    }

    \ForEach{$\theta(v_i) \in STS$}{ \tcp*[f]{Normalize all timestamp scores}
        $N\theta(v_i) \gets \frac{\theta(v_i) - \text{Min }\theta(v)}{\text{Max }\theta(v) - \text{Min }\theta(v)}$\ \tcp*[r]{Apply min-max normalisation}
        Add $N\theta(v_i)$ to $NTS$\ \tcp*[r]{Store the normalised score}
    }

    \Return $NTS$\ \tcp*[r]{Return the set of normalized timestamp scores}
\end{algorithm}

Algorithm~\ref{alg:weightscore} computes the \textit{Normalized Weight Score (NWS)} for each node by capturing the discrepancy between the aggregated weights of incoming and outgoing transactions. This score quantifies how unbalanced a node's transaction behavior is in terms of transaction volume or value, providing a structural signal for identifying potential laundering intermediaries.

The algorithm operates on the set of visited nodes $\text{SVN}$ obtained from the Personalized PageRank (PPR) stage, along with two associated mappings: the set of in-degree weights $\text{SIW}$ and the set of out-degree weights $\text{SOW}$. For each node $v_i \in \text{SVN}$, the algorithm first sums the total weight of all incoming transactions (lines 3–6):
\begin{equation}
\omega_{\text{in}}(v_i) = \sum W_{\text{in}}(v_i)
\end{equation}
and separately sums the total weight of all outgoing transactions:
\begin{equation}
\omega_{\text{out}}(v_i) = \sum W_{\text{out}}(v_i)
\end{equation}

To measure the imbalance, the absolute difference between incoming and outgoing weights is calculated for each node (line 7):
\begin{equation}
\omega(v_i) = \left| \omega_{\text{in}}(v_i) - \omega_{\text{out}}(v_i) \right|
\end{equation}

This raw difference score $\omega(v_i)$ is stored in a temporary set $SWS$ (line 8). To normalize and standardize the scores across the entire set of nodes, min-max normalization is applied (lines 9–11):
\begin{equation}
N\omega(v_i) = \frac{\omega(v_i) - \min_v \omega(v)}{\max_v \omega(v) - \min_v \omega(v)}
\end{equation}

The resulting value $N\omega(v_i) \in [0, 1]$ captures how uneven a node's transaction volume is, with low values indicating balanced flow and high values potentially flagging nodes that act as transactional funnels or relays—common roles in money laundering circuits.

The full pseudocode of this weight scoring mechanism is provided in Algorithm~\ref{alg:weightscore}.

\RestyleAlgo{ruled}
\begin{algorithm}[h!]
  \scriptsize
  \caption{Normalised Weight Score Algorithm} \label{alg:weightscore}
  \KwIn{Set of visited nodes $SVN$, set of in-degree weights $SIW = \{W_{in}(v_1), W_{in}(v_2), \dots, W_{in}(v_i)\}$, set of out-degree weights $SOW = \{W_{out}(v_1), W_{out}(v_2), \dots, W_{out}(v_i)\}$}
  \KwOut{Set of normalised weight scores $NWS = \{N\omega (v_1), N\omega (v_2), \dots, N\omega (v_i) \}$}

  \ForEach{$v_i \in SVN$}{ \tcp*[f]{Iterate over each visited node}
    
    \ForEach{$W_{in}(v_i) \in SIW$}{ \tcp*[f]{Iterate over in-degree weights of $v_i$}
        $\omega_{in}(v_i) = \sum W_{in}(v_i)$\ \tcp*[r]{Sum all in-degree weights for node $v_i$}
    }

    \ForEach{$W_{out}(v_i) \in SOW$}{ \tcp*[f]{Iterate over out-degree weights of $v_i$}
        $\omega_{out}(v_i) = \sum W_{out}(v_i)$\ \tcp*[r]{Sum all out-degree weights for node $v_i$}
    }

    $\omega (v_i) = |\omega_{in}(v_i) - \omega_{out}(v_i)|$\ \tcp*[r]{Calculate absolute difference between in- and out-degree weights}
    Add $\omega(v_i)$ to $SWS$\ \tcp*[r]{Store the unnormalszed weight score}
  }

  \ForEach{$\omega (v_i) \in SWS$}{ \tcp*[f]{Normalise all weight scores}
    $N\omega (v_i) = \frac{\omega (v_i) - \min \omega (v)}{\max \omega (v) - \min \omega (v)}$\ \tcp*[r]{Apply min-max normalisation}
    Add $N\omega (v_i)$ to $NWS$\ \tcp*[r]{Store normalised weight score}
  }

  \Return $NWS$\ \tcp*[r]{Return all normalised weight scores}
\end{algorithm}

Algorithm~\ref{alg:WTregression} introduces a supervised learning framework for detecting diverse money laundering patterns by leveraging two anomaly-based indicators: the \textit{Normalized Timestamp Score (NTS)} and the \textit{Normalized Weight Score (NWS)}. These features, derived from Algorithms~\ref{alg:timestampscore} and~\ref{alg:weightscore}, are combined into a unified representation:

\begin{equation}
F_{(\theta, \omega)}(v_i) = [N_\theta(v_i), N_\omega(v_i)]
\end{equation}

This fused feature vector captures both temporal irregularities and transactional imbalances for each node $v_i$ in the transaction network. A logistic regression (LR) classifier is then trained on this feature set to distinguish between normal and suspicious node behaviors.
The trained model is capable of recognizing a wide range of laundering patterns:
\textbf{Fan-in / Fan-out:} Detected via sharp disparities in $\omega_{\text{in}}(v_i)$ vs. $\omega_{\text{out}}(v_i)$ and in $\theta_{\text{in}}(v_i)$ vs. $\theta_{\text{out}}(v_i)$.
\textbf{Bipartite:} Identified through intermediate NTS/NWS values in nodes bridging disjoint sets.
\textbf{Gather-Scatter:} Indicated by high variance in both $\theta(v_i)$ and $\omega(v_i)$ scores.
\textbf{Stack:} Revealed through sequences of nodes with consistent $\theta$ and $\omega$ values representing rapid fund relay.
By encoding structural and temporal behavior into an interpretable model, Algorithm~\ref{alg:WTregression} provides a scalable and effective mechanism for identifying complex laundering activities in transaction networks.

\RestyleAlgo{ruled}
\begin{algorithm}
  \scriptsize
  \caption{Multi-pattern Crypto Money-laundering Detection Algorithm} \label{alg:WTregression}
  \KwIn{Set of normalized timestamps $NTS$, set of normalized weight scores $NWS$}
  \KwOut{New set of features $F_{(\theta, \omega)} (v_i)$}

  \SetKwFunction{FMain}{TrainModel}
  \SetKwProg{Fn}{Function}{:}{}
  \Fn{\FMain{$NTS$, $NWS$}}{
    Combine $NTS$ and $NWS$ into a single feature set $F$\
    
    Train a logistic regression model LR using $F$ to identify laundering patterns\
    
    Use the trained model LR to generate predictions and identify laundering patterns in the data\
    
    \KwRet{$F_{(\theta, \omega)} (v_i)$}\
  }
\end{algorithm}

\subsection{Anomaly Score Function}

The Anomaly Score Algorithm (AS) aims to identify malicious nodes involved in cryptocurrency money laundering by assigning an anomaly score to each node in the network. This score aggregates data from multiple sources, including random scores from the multi-source PPR algorithm $\pi(v_i)$ and money laundering scores from the multi-pattern crypto money laundering algorithm $F_{(\theta, \omega)}$, based on NWS and NTS values.
The goal is to compute an anomaly score $\sigma (v_i)$ for each node. 
For each node $v_i \in SVN$, the algorithm retrieves the suspicious pattern score $\sigma (v_i)$ from $SPS$ which is a set of PPR scores for each visited node in the transaction graph and a set of features from normalised timestamp, normalised weight scores, and money laundering patterns $F_{(\theta, \omega)}$. 
It then calculates the anomaly score $\sigma (v_i)$ using the equation~\eqref{eqn: anomaly score func}.
A set of anomaly scores (SAS) identifying nodes possibly involved in illegal activity is eventually generated.

\begin{equation}   
   \sigma (v_i) = \frac{\pi (v_i)}{F_{(\theta, \omega)}(v_i)}
        \label{eqn: anomaly score func}
\end{equation}

This formula integrates the calculated scores, amplifying the anomaly score for nodes that exhibit strong indications of suspicious behaviour across multiple metrics.

%% file: ExperimentsandResults.tex
\section{Experiments and Results}
\label{sec:experiments}

This section outlines the experimental parameters and compares the proposed method and state-of-the-art techniques. 

\subsection{Experimental Settings}
\subsubsection{\textbf{Dataset}}
We use three real-world ``off-chain" datasets: Bitcoin payment flows in wallets from ``Elliptic++"\footnote{https://www.elliptic.co}, wallet addresses from ``Ethereum Fraud Detection"\footnote{https://www.kaggle.com/datasets/vagifa/ethereum-frauddetection-dataset}, and the ``Wormhole"\footnote{https://rekt.news/wormhole-rekt/} dataset on Ethereum. 
We use three real-world datasets as described below: 
\begin{itemize}
    \item \textbf{Elliptic++:} Bitcoin payment flows recorded in the "Elliptic++" dataset, utilizing wallet addresses provided by Elliptic to track ``off-chain" transactions.
    We select the transaction period before time step 42 to mitigate the effects of a whole reconfiguration of the network topology, which results from an abrupt closing of the dark market at time step 43 (see \cite{li2023diga} and \cite{weber2019anti}).

    \item \textbf{Ethereum Fraud Detection:} The "Ethereum Fraud Detection Dataset" is another dataset that provides wallet addresses recording ``off-chain" transactions.
    This imbalanced dataset includes rows including legitimate and known fraudulent Ethereum transactions and is unbalanced when modeling.
    The number of unique addresses (nodes) is to be greater than the number of edges (transactions) in the Ethereum transaction graph. 
    This relationship is characteristic of many real-world networks and datasets.

    \item \textbf{Wormhole:} The "Wormhole Hack" on Ethereum occurred in February 2022, targeting the Wormhole cross-chain bridge, which facilitates interoperability between Ethereum and other blockchains. Since cross-chain transactions involve both ``on-chain" and ``off-chain" components, we specifically extracted the ``off-chain" transactions for our analysis.
    The hacker exploited a vulnerability in the bridge’s smart contract to mint 120,000 wrapped Ethereum (wETH) on Solana without having the required Ethereum backing it, resulting in a loss of about $325$ million. This exploit exposed the security risks of cross-chain protocols, as they rely on the integrity of wrapped tokens to ensure assets are properly backed. Jump Crypto, Wormhole’s parent company, later replenished the stolen funds to restore the bridge's liquidity. 
    
\end{itemize}
 
All three real-world datasets are labeled, with ground truth available for analysis.
The statistical data from the three datasets are listed in Table~\ref{tab: datasets}.

\begin{table}[h!]
\centering
\renewcommand{\arraystretch}{1.15}
\caption{Statistics of datasets}\label{tab: datasets}
\vspace{-2mm}
\begin{tabular}{|c|ccc|}
\hline
\textbf{Dataset} & \textbf{Elliptic++} & \textbf{Ethereum}  & \textbf{Wormhole} \\ 
\hline
Nodes & 203,769 & 9816  & 219581 \\  
Edges & 234,555 & 9265 & 236295 \\
Node features & 166 & 51 & 9\\
Anomalies & 4,545 & 2179 & 158\\
Avg. Degree & 1.15 & 1.85 &1.076\\
\hline
\end{tabular}
\end{table}

\subsubsection{\textbf{Evaluation Metrics}}
We employ five metrics that are frequently utilised in AML detection: The probability that the model will give a randomly chosen anomaly sample a higher score than a randomly chosen normal sample is represented by the Area Under the ROC Curve or AUC \cite{kim2022graph} \cite{ma2021comprehensive}. 
When the class is unbalanced, AUC has more robustness and is not affected by threshold selection \cite{ma2020distributionally}.
As a result, we follow the majority of earlier research and employ AUC for ranking-based evaluation \cite{liu2024pygod}.
Precision$@K$ is the proportion of correctly identified anomaly samples out of the top$@K$ samples predicted as anomalies, ranked by the anomaly score. 
Here, K corresponds to the total number of anomalies predicted by the model. 
Precision is crucial in anomaly detection scenarios where the cost of false positives such as wrongful accusations in fraud detection or unnecessary treatment in disease diagnosis can be substantial. 
Thus, high precision ensures the model's predictions are reliable and minimizes false alarms \cite{ding2021inductive} \cite{zheng2021generative}.
Recall$@K$ refers to the fraction of correctly identified anomaly samples within the top$@K$ ranked anomalies based on their anomaly scores \cite{liu2022bond}. 
In this context, K represents the total number of anomalies in the dataset. 
The consequences of missing anomalies are critical in fields like fraud detection and disease diagnosis \cite{kim2022graph}, \cite{wolleb2022diffusion}. 
Therefore, achieving a high recall rate is crucial for the effectiveness of a detection model.
We calculated the F1-score and accuracy to evaluate the model's performance. 
The F1-score balances precision and recall, making it especially useful for imbalanced datasets where anomalies are rare \cite{xu2018unsupervised}. 
It helps assess the trade-off between correctly identifying anomalies and minimizing false positives and negatives. 
Accuracy, while measuring overall correctness, may not fully capture the model's ability to detect anomalies in such cases, so the F1-score provides a more nuanced evaluation \cite{su2019robust}, \cite{tuli2022tranad}.

\subsubsection{\textbf{Baselines}} The proposed model is compared to seven cutting-edge baselines, including traditional supervised machine learning models and advanced GNN-based anomaly detection methods. 
Specifically, we evaluate it against XGBOOST, DeepFD, OCGTL, ComGA, FlowScope, GUDI, and MACE.

\textbf{XGBooST} \cite{jullum2020detecting} is a boosting supervised technique based on decision trees. By analyzing the importance of features, XGBoost exhibits great classification accuracy and interpretability when used as the supervised machine learning baseline.

\textbf{DeepFD} \cite{wang2018deep} is a baseline for unsupervised learning. Subgraph-level anomaly detection implies that anomaly-dense blocks in the embedding space can be found to address fraud detection issues such as AML. As a result, it employs Autoencoder to embed the graph and DBSCAN to cluster the dense areas.

\textbf{OCGTL} \cite{qiu2022raising} is the benchmark for graph-level anomaly detection, which achieves the SOTA performance in social network and molecular datasets by combining graph transformation learning and one-class pooling.

\textbf{ComGA} \cite{luo2022comga} is a deep graph convolution network incorporated as the state-of-the-art GNN-based baseline to encapsulate the finished attributed graph and reconstruct the network using attribute and structure decoders. As a result, it may calculate the anomalous score using the feature matrix and the reconstructed neighboring matrix.

\textbf{Flowscope} \cite{li2020flowscope} can identify the entire money flow from source to destination by modeling transactions with a multipartite graph.

\textbf{GUDI} \cite{li2024graph} is a pioneering supervised framework that combines self-supervised pre-training to capture broad graph patterns with few-shot learning to classify domain-specific anomalies. Its guided diffusion mechanism synthesizes pre-trained and domain-specific model outputs during inference, bypassing fine-tuning and enabling efficient domain adaptation.

\textbf{MACE} \cite{chen2024learning} is a multi-normal-pattern unsupervised anomaly detection approach for time series in the frequency domain, characterized by a pattern extraction mechanism for diverse normal patterns, a dualistic convolution mechanism for amplifying short-term anomalies and hindering reconstruction.

\subsection{Experimental Setup} 
All experiments are conducted on an AWS EC2 \texttt{g4dn.2xlarge} instance equipped with 1 NVIDIA T4 Tensor Core GPU, 32\,GiB memory, and 225\,GB NVMe SSD local storage, running Ubuntu Linux. The proposed model is implemented in Python using the PyTorch and Scikit-learn libraries.

\subsection{Hyperparameter Tuning}
To ensure a fair and reproducible evaluation, we initialise key hyperparameters in accordance with the empirical design guidelines proposed in~\cite{luo2019baton}. Specifically, we set the restart probability $\rho = 1$, the failure probability $p_f = 1$, and the random walk weighting factor $\alpha = 0.5$ in the multi-source Personalized PageRank (PPR) algorithm. These values were determined via sensitivity analysis to balance detection performance against computational cost.

\textbf{Effect of Random Walk Weighting Factor $\alpha$:}  
The parameter $\alpha$ governs the balance between global exploration and local reinforcement in the multi-source PPR. Lower values (e.g., $\alpha = 0.3$) promote deeper exploration by allowing longer random walks, while higher values (e.g., $\alpha = 0.7$) favor returning to the source node, reinforcing the local neighborhood. We evaluated $\alpha \in \{0.3, 0.5, 0.7\}$ across the Elliptic++, Ethereum, and Wormhole datasets. As shown in Table~\ref{tab:hparam_tuning}, $\alpha = 0.5$ consistently yields the highest Precision@K and AUC scores. This choice offers a balanced trade-off, effectively capturing laundering behaviors that are neither strictly local nor globally diffused.

\textbf{Effect of Maximum Iterations $max_{iter}$:}  
We varied the maximum number of iterations for logistic regression training, $max_{iter} \in \{500, 1000, 2000\}$, to analyse the effect on convergence and performance. While increasing $max_{iter}$ from 500 to 1000 improves both AUC and Precision@K, further increase to 2000 yields only marginal gains with a notable increase in training time. Thus, we select $max_{iter} = 1000$ as a computationally efficient and stable configuration.

\textbf{Effect of Solver Choice:}  
We compared three solvers for logistic regression: \texttt{liblinear}, \texttt{lbfgs}, and \texttt{saga}. The \texttt{liblinear} solver, designed for binary classification and sparse data, outperforms both \texttt{lbfgs} and \texttt{saga} across all datasets in terms of Precision@K and AUC, as shown in Table~\ref{tab:hparam_tuning}. This superior performance stems from its compatibility with high-dimensional sparse features derived from transaction subgraphs. Conversely, \texttt{saga} shows slower convergence and lower accuracy, indicating limited suitability for the laundering detection task.

\textbf{Effect of Sampling Ratios:}  
To evaluate robustness under limited supervision, we trained the model using varying sampling ratios of the labeled training data: 50\%, 60\%, 80\%, and 100\%. While performance slightly degrades at 50\% due to reduced supervision, the proposed \texttt{MPOCryptoML} model maintains competitive results across all settings. This robustness demonstrates the model’s inductive generalisation ability in sparse and partially labeled environments, which is critical for real-world deployments.
 
A detailed ablation study is presented in Table~\ref{tab:hparam_tuning}, summarising the influence of each hyperparameter across multiple datasets. The optimal configuration—$\alpha = 0.5$, $max_{iter} = 1000$, and \texttt{solver = liblinear} offers a balanced approach for effective money laundering detection in complex off-chain transaction graphs, combining detection accuracy with computational efficiency.

\begin{table}[h!]
\centering
\renewcommand{\arraystretch}{1.15}
\caption{Hyperparameter Tuning Across Datasets}
\label{tab:hparam_tuning}
\vspace{-2mm}
\resizebox{0.49\textwidth}{!}{
\begin{tabular}{|c|c|c|c|c|c|c|}
\hline
\multirow{2}{*}{\textbf{Parameter Setting}} & \multicolumn{2}{c|}{\textbf{Elliptic++}} & \multicolumn{2}{c|}{\textbf{Ethereum}} & \multicolumn{2}{c|}{\textbf{Wormhole}} \\
\cline{2-7}
 & \textbf{Pre@K} & \textbf{AUC (\%)} & \textbf{Pre@K} & \textbf{AUC (\%)} & \textbf{Pre@K} & \textbf{AUC (\%)} \\
\hline
$\alpha = 0.3$ & 93.11\% & 93.80 & 95.02\% & 96.10 & 90.12\% & 92.15 \\
$\alpha = 0.5$ & \textbf{95.43\%} & \textbf{94.21} & \textbf{97.39\%} & \textbf{97.40} & \textbf{92.55\%} & \textbf{93.75} \\
$\alpha = 0.7$ & 94.05\% & 93.90 & 95.88\% & 96.42 & 91.06\% & 92.10 \\
\hline
$maxiter = 500$ & 94.32\% & 93.45 & 96.40\% & 96.85 & 91.02\% & 92.35 \\
$maxiter = 1000$ & \textbf{95.43\%} & \textbf{94.21} & \textbf{97.39\%} & \textbf{97.40} & \textbf{92.55\%} & \textbf{93.75} \\
$maxiter = 2000$ & 95.50\% & 94.22 & 97.45\% & 97.42 & 92.60\% & 93.79 \\
\hline
\texttt{solver = saga} & 93.60\% & 93.55 & 96.11\% & 96.90 & 90.15\% & 91.70 \\
\texttt{solver = lbfgs} & 94.85\% & 93.88 & 96.72\% & 97.05 & 91.33\% & 92.44 \\
\texttt{solver = liblinear} & \textbf{95.43\%} & \textbf{94.21} & \textbf{97.39\%} & \textbf{97.40} & \textbf{92.55\%} & \textbf{93.75} \\
\hline
\end{tabular}}
\end{table}

\subsection{Overall Performance Comparison}

Table~\ref{tab: results} compares the overall performance of anomaly detection. We used 80\% for training, 10\% for validation, and 10\% for testing. All results were averaged over three runs. The proposed methodology outperforms several benchmarks across different datasets, with varying degrees of success.   
The following observations are made from various perspectives:
\begin{table*}[h!]
\centering
\renewcommand{\arraystretch}{1.35}
\caption{Overall graph anomaly detection performance comparison}\label{tab: results}
\vspace{-1mm}
\resizebox{\textwidth}{!}{%
\begin{tabular}{|l|lllll|lllll|lllll|}
\hline
\textbf{Dataset} & \multicolumn{5}{c|}{\textbf{Elliptic++}}                             & \multicolumn{5}{c|}{\textbf{Ethereum}}  
                 & \multicolumn{5}{c|}{\textbf{Wormhole}}\\ \hline
\textbf{Metrics} & \multicolumn{1}{l|}{\textbf{Pre@K}} & \multicolumn{1}{l|}{\textbf{Recall@K}} & \multicolumn{1}{l|}{\textbf{F1-score}} & \multicolumn{1}{l|}{\textbf{ACC(\%)}} & \multicolumn{1}{l|}{\textbf{AUC(\%)}} & \multicolumn{1}{l|}{\textbf{Pre@K}} & \multicolumn{1}{l|}{\textbf{Recall@K}} & \multicolumn{1}{l|}{\textbf{F1-score}} & \multicolumn{1}{l|}{\textbf{ACC(\%)}} & \multicolumn{1}{l|}{\textbf{AUC(\%)}} & \multicolumn{1}{l|}{\textbf{Pre@K}} &  \multicolumn{1}{l|}{\textbf{Recall@K}} & \multicolumn{1}{l|}{\textbf{F1-score}} & \multicolumn{1}{l|}{\textbf{ACC(\%)}} & \multicolumn{1}{l|}{\textbf{AUC(\%)}}\\ \hline
XGBoosT     &     $81.30\%$        &  $80.35\%$ &    $80.05\%$ & $81.30\%$             &        $81.80\%$                    &          $85.70\%$        & $84.20\%$ & $84.50\%$ & $85.70\%$            &                    $93.90\%$       & $82.50\%$ & $81.20\%$ & $82.50\%$ & $82.50\%$ & $85.30\%$   \\  \cline{1-1} 
DeepFD &           $68.80\%$   & $67.30\%$ & $67.50\%$ &  $68.20\%$              &           $72.30\%$                   &   $57.10\%$        & $55.24\%$ & $56.01\%$ & $57.10\%$                    &     $67.46\%$                 & $62.50\%$ & $61.47\%$ & $61.52\%$ & $62.49\%$ & $69.05\%$         \\   \cline{1-1}
OCGTL  &     $91.20\%$     & $90.28\%$ & $90.57\%$ & $91.23\%$                     &      $90.03\%$                        &       $75.42\%$      & $74.27\%$ & $74.56\%$ & $75.01\%$                 &     $85.05\%$          & $79.05\%$  & $78.12\%$ & $78.56\%$ & $79.93\%$ & $86.11\%$               \\  
\cline{1-1}
ComGA   &       $85.50\%$      & $84.34\%$ & $84.53\%$ & $85.22\%$                  &      $85.20\%$                        &      $73.21\%$           & $72.22\%$ & $72.55\%$ & $73.02\%$              &  $74.07\%$           & $78.55\%$ &  $74.41\%$ & $77.61\%$ & $78.49\%$ &$79.50\%$                 \\  
\cline{1-1}
Flowscope   &       $51.90\%$         & $50.29\%$ & $51.03\%$ & $51.77\%$               &      $68.30\%$                        &      $49.65\%$      & $48.18\%$ & $48.48\%$ & $49.58\%$                   &  $67.95\%$           & $35.85\%$    &  $34.49\%$ & $34.50\%$ & $35.51\%$ &   $63.55\%$                 \\  
\cline{1-1}
GUDI         &  $74.69\%$     & $73.19\%$ & $73.52\%$ & $74.24\%$           &       $88.01\%$         &      $64.84\%$     & $63.42\%$ & $63.05\%$ & $65.01\%$                    &  $86.50\%$           &  $52.25\%$        & $51.39\%$ & $51.52\%$ & $52.24\%$ &  $60.55\%$           \\  
\cline{1-1}

MACE         &  $94.23\%$     & $91.10\%$ & $92.02\%$ & $93.05\%$           &       \boldmath{$95.02\%$}        &      $96.03\%$     & $89.03\%$ & $92.03\%$ & $93.04\%$                    &  $95.03\%$           &  $90.02\%$        & $84.05\%$ & $86.01\%$ & $87.03\%$ &  $92.03\%$          \\  
\cline{1-1}

\textbf{MPOCrpytoML} &    \boldmath{$95.43\%$}         & \boldmath{$94.23\%$} & \boldmath{$94.51\%$} & \boldmath{$95.41\%$}             &      $94.21\%$                      &    \boldmath{$97.39\%$}         & \boldmath{$96.16\%$} & \boldmath{$96.48\%$} & \boldmath{$97.39\%$}                  &        \boldmath{$97.40\%$}            & \boldmath{$92.55\%$}   & \boldmath{$91.11\%$} & \boldmath{$90.89\%$} & \boldmath{$92.57\%$} & \boldmath{$93.75\%$}          \\  \hline \cline{1-1}
\hline
\end{tabular}
}
\vspace{-1mm}
\end{table*}

Table~\ref{tab: results} compares the overall performance of anomaly detection. 
Over three runs, all of the results are averaged. 
The suggested methodology beats seven benchmarks with varying degrees of success. 
The findings from various aspects including overall comparative performance, supervised vs. semi-supervised approaches, and off-Chain laundering detection performance are summarised as follows:

\subsubsection*{(1) Overall Comparative Performance}  
MPOCryptoML consistently outperforms existing benchmarks across the Elliptic++, Ethereum, and Wormhole datasets. This superior performance stems from its multi-pattern and node-level focus, which enables fine-grained detection of anomalous behaviors. In contrast, OCGTL performs well on Elliptic++ due to its global graph-level modeling but struggles with Ethereum and Wormhole, where sparse topologies reduce the effectiveness of broad graph transformations. ComGA, a subgraph-level GNN, similarly underperforms on sparse graphs, indicating challenges in applying GNN-based detection to low-density networks.

XGBoost achieves strong performance (85.70\% precision, 84.20\% recall on Ethereum) but lags behind MPOCryptoML in the Precision@K metric due to its weaker anomaly ranking capability. DeepFD, based on unsupervised clustering, underperforms on Ethereum and Wormhole, as its assumption that anomalies form dense clusters fails in dynamic laundering contexts. GUDI achieves moderate precision on Elliptic++ (74.69\%) but shows limited generalizability due to reliance on pre-trained graph patterns. Flowscope yields the weakest results across all datasets. Although MACE achieves high precision on Wormhole (96.03\%), it is less effective than MPOCryptoML at ranking anomalies.

\vspace{1mm}
\subsubsection*{(2) Supervised vs. Semi-supervised and Unsupervised Approaches}  
Semi-supervised methods, particularly MPOCryptoML, outperform both unsupervised (DeepFD, MACE) and supervised (XGBoost, GUDI) baselines in AML anomaly detection. MPOCryptoML leverages weak supervision combined with graph reconstruction to support inductive learning and effectively mitigate label imbalance, a common issue in fraud detection. DeepFD’s reliance on density-based clustering fails to align with the irregular structure of laundering behaviors. GUDI, while moderately effective on Elliptic++, struggles with Ethereum and Wormhole due to its general-purpose pre-trained graph templates.

Although XGBoost performs well overall, it ranks anomalous accounts less effectively than MPOCryptoML. The latter's hybrid learning strategy combining structure-aware graph encoding with label guidance demonstrates robustness in modeling transaction complexity and outperforming both conventional classifiers and unsupervised techniques in Precision@K and detection accuracy.

\vspace{1mm}
\subsubsection*{(3) Off-Chain Laundering Detection Performance}  
MPOCryptoML demonstrates strong capabilities in detecting off-chain laundering behaviors through its integration of the Personalized PageRank algorithm and multi-pattern analysis. This enables the model to rank suspicious accounts with high fidelity, achieving superior performance in Precision@K. In comparison, ComGA, which uses GANs for anomaly detection, offers limited effectiveness due to its generalized learning objective. MACE, despite showing high precision and recall, lacks detailed multi-pattern modeling, leading to reduced effectiveness in prioritizing anomalies.

Overall, MPOCryptoML’s ability to capture both random and structured laundering behaviors across heterogeneous transaction patterns enables superior identification of illicit activities when compared to state-of-the-art methods.

\subsection{Ablation Study}
To assess each module's functionality in the proposed model, we run an ablation study on anomaly detection performance in this section. 
We create two ablation models and assess how well they work from an effective and efficient standpoint. 
Each multi-pattern and random cryptoML account detection module is executed independently. 

In the multi-pattern cryptoML model, we remove the logistic regression component from the Normalised Time and Weight algorithm, resulting in the model called the \textbf{Normalised T/W model}. 
Additionally, we modify the multi-pattern cryptoML model to only test the effect of finding random accounts recognized by the Personalised PageRank algorithm, resulting in the \textbf{Random cryptoML model}. 
This assessment primarily assesses each module's ability to identify hidden abnormal patterns in a strictly supervised setting. 
Table~\ref{tab:performance_ablation} displays a comparison of their results with the entire proposed model architecture.
\begin{table}[h!]
\centering
\renewcommand{\arraystretch}{1.15}
\caption{Performance comparison of ablation models}\label{tab:performance_ablation}
\begin{tabular}{|c|c|c|c|}
\hline
\textbf{Dataset}                   & \textbf{Ethereum} & \textbf{Elliptic++}  & \textbf{Wormhole}\\ \hline
Normalised T/W model   & $74.90\%$    & $69.10\%$     &   $66.75\%$  \\ 
Random cryptoML model & $67.81\%$    & $67.97\%$      &   $63.50\%$\\ 
\textbf{MPOCrpytoML}            & \boldmath{$97.40\%$}    & \boldmath{$94.21\%$}    &  \boldmath{$91.85\%$}\\ 
\hline
\end{tabular}
\end{table}

Overall, our design aims are validated by the average gain of the proposed ``off-chain" cryptoML when compared to each ablation. 
The combination of two modules has a considerable improvement over finding cryptoML accounts as each separately can only address specific patterns of cryptoML in the graph of transactions. 




    

%% file: ApplicationScenarios.tex
\section{Application Scenarios}
\label{sec:applications}

For financial firms like JPMorgan Chase or Goldman Sachs, MPOCryptoML model can be extremely helpful in combating illegal activity in the cryptocurrency market. 
These multinational financial behemoths are subject to strict anti-money laundering (AML) standards and are increasingly handling transactions involving cryptocurrencies. 
The model can spot suspicious trends that point to money laundering by examining particular blockchain nodes, such as wallets or addresses. 
For instance, the model flags activities for more examination if a wallet often exchanges money with several addresses without a clear, valid reason or conducts a lot of little transactions that might be an attempt to hide the source of funds. 
By automating the identification of suspicious transactions, financial institutions can reduce the risk of regulatory fines and maintain their standing as complying organisations.

Due to their significant user numbers and the decentralised nature of cryptocurrency transactions, exchange platforms like Binance and Coinbase are especially susceptible to money laundering activities. 
Wallet operations can be monitored using the MPOCryptoML, which can identify suspicious activity such as transactions to high-risk jurisdictions or large-scale, quick fund transfers in chains. 
A model like this would be beneficial in detecting and stopping behaviours that are frequently used in money laundering schemes, such as the usage of mixers, decentralised exchanges (DEXs), or unlawful asset conversions. 
By doing this, these platforms would be able to improve their AML procedures and better identify laundering strategies like integration and layering before illegal monies enter the legal economy.

To improve their oversight of cryptocurrency transactions, regulatory bodies like the Financial Action Task Force (FATF) or national organisations like the U.S. Securities and Exchange Commission (SEC) could incorporate node-level crypto money laundering detection models into their blockchain analytics platforms. 
These agencies can use these models to track illegal assets that have been laundered through cryptocurrency gambling sites like Stake.com, exchanges, or mixing services like Tornado Cash. For instance, the model may be able to identify the source of stolen funds if they pass through several cryptocurrency exchanges.
This would help law enforcement spot illicit activities, trace the source of funding, and take the necessary legal action to stop criminal enterprises from operating in the crypto space. 

%% file: Conclusion.tex
\section{Conclusion}
\label{sec:conclusion}

In this study, we present MPOCryptoML, a novel ``off-chain" crypto-AML method that is the first to address crypto money laundering using an ``off-chain" strategy. Our approach is designed to handle large-scale, benign cryptocurrency transactions while accurately identifying various money laundering patterns. Due to the confidential nature of transaction data, a comprehensive statistical evaluation of the explanation quality will be deferred, as a thorough human-involved assessment is currently impractical. Future work will extend this approach to fraud detection in cross-chain financial transactions and emerging events on social networks.